\documentclass[12pt,a4paper]{article}

\usepackage[utf8]{inputenc}
\usepackage[english]{babel}
\usepackage[T1]{fontenc}
\usepackage{amsmath,amssymb,amsfonts,mathtools}
\usepackage{graphicx}
\usepackage{booktabs}
\usepackage{array,multirow}
\usepackage{float}
\usepackage{caption}
\usepackage{subcaption}
\usepackage{hyperref}
\usepackage{url}
\usepackage{xcolor}
\usepackage{geometry}
\usepackage{enumitem}
\usepackage{fancyhdr}
\usepackage[numbers,sort&compress]{natbib}
\title{Multi-Channel Di-Higgs Production in the scNMSSM: From Exotic $8b$ to Boosted $8\tau$ Signatures at the HL-LHC}
\author{Marwa Telba}
\date{}

\begin{document}
\maketitle

\noindent\textbf{Keywords:} 
di-Higgs production; semi-constrained NMSSM; 
High-Luminosity LHC; exotic Higgs decays; 
multi-$b$ final states; boosted multi-tau topology; Higgs self-coupling; light pseudoscalar

\section*{abstract}
We study the signal topologies of Higgs-pair production in the semi-constrained NMSSM at large $\lambda$ and low $\tan\beta$ at the HL-LHC ($\sqrt{s}=13.6$~TeV, $\mathcal{L}=3000$~fb$^{-1}$), using four benchmark points drawn from a parameter scan in which the LHC Higgs-sector constraints were not applied. The benchmarks cover exotic $h_1h_1\to4a_1\to8b$ (BP1, BP3), boosted $h_1h_1\to4a_1\to8\tau$ (BP2) and SM-like (BP4) final states. Signals are simulated with \textsc{MadGraph5\_aMC@NLO}, \textsc{Pythia}~8.316 and \textsc{Delphes}~3, with $10^5$ events per benchmark. Background yields are taken from published analyses and serve only as illustrative estimates.
In this revised version (v2) we add an a posteriori check with \textsc{NMSSMTools}. All four benchmark points are flagged: BP1--BP3 have $\mathrm{BR}(h_1\to a_1a_1)>99\%$, in conflict with the limits on undetected Higgs decays and with CMS searches for $h\to aa\to4\tau$ and $4b$, and BP4 deviates from the measured $b$- and $\tau$-couplings by more than $2\sigma$. The benchmarks therefore do not represent experimentally allowed parameter points, the signal yields apply to the stated branching fractions only, and the earlier statement of $5\sigma$ sensitivity is withdrawn. The signal rate scales as $\mathrm{BR}(h\to aa)^2$, so a branching fraction compatible with current limits would reduce the yields by one to two orders of magnitude.
The detector-level characterisation of the topologies remains of methodological interest. For BP2 ($m_{a_1}=8.2$~GeV), 82.4\% of same-$a_1$ $\tau$ pairs satisfy $\Delta R(\tau,\tau)<0.4$ owing to the boost $\gamma\approx7.5$, which motivates dedicated boosted-$a_1$ tagging.

\noindent\textbf{Note added in v2.} The first version of this paper described the benchmark points as viable and quoted an indicative $5\sigma$ sensitivity for the $8b$ channel. Those statements did not take into account the experimental limits on exotic Higgs decays, and they are corrected here (see Section~\ref{subsec:lhc_check}).

\section{Introduction}
\label{sec:introduction}

The observation of the Higgs boson at the Large Hadron Collider (LHC) in 2012~\cite{ATLAS:2012yve,CMS:2012qbp} completed the particle content of the 
Standard Model (SM), yet left open a fundamental question: what is the precise shape of the Higgs potential? The answer is encoded in the Higgs self-coupling 
$\lambda_{hhh}$, which in the SM is uniquely determined by the Higgs boson mass and vacuum expectation value as $\lambda_{hhh}^{\rm SM} = 3m_h^2/v$. 
Direct experimental access to this coupling requires the measurement of Higgs boson pair (di-Higgs) production, a process that proceeds predominantly 
via gluon fusion at the LHC through a destructive interference between top-quark triangle and box diagrams~\cite{Borowka:2016ehy,deFlorian:2013jea}. The resulting 
SM cross section of $\sigma_{\rm SM} \approx 33$~fb at $\sqrt{s} = 13.6$~TeV makes this one of the rarest processes accessible at the LHC, and its measurement is 
among the primary physics goals of the High-Luminosity LHC (HL-LHC)~\cite{Cepeda:2019klc,
LHCHiggsCrossSectionWorkingGroup:2016ypw}.

In theories beyond the Standard Model (BSM), the Higgs self-coupling can deviate substantially from its SM value, modifying both the di-Higgs production rate and the kinematic distributions of the final-state particles. Such deviations are particularly well-motivated in supersymmetric extensions of the SM, where the enlarged Higgs sector introduces new scalar states and additional contributions to the trilinear Higgs coupling. In the Next-to-Minimal Supersymmetric Standard Model (NMSSM), the Higgs sector is further extended by a gauge-singlet superfield, giving rise to a rich phenomenology that includes a light CP-odd pseudoscalar $a_1$ and significant modifications of the Higgs self-coupling relative to the SM~\cite{Ellwanger:2009dp,Wu:2015, Ellwanger:2013ova,Cao:2013si}.

A particularly important regime of the NMSSM is that of large singlet-doublet 
coupling $\lambda$ and low $\tan\beta$, which is theoretically motivated by 
considerations of naturalness and electroweak baryogenesis. In this regime, the lightest CP-even Higgs boson $h_1$ can be identified with the observed 125~GeV state while 
simultaneously accommodating a light pseudoscalar $a_1$ with mass $m_{a_1}$ below the $b\bar{b}$ threshold. When kinematically accessible, the exotic decay $h_1 \to a_1 a_1$ can dominate the $h_1$ width, giving rise to multi-particle final states qualitatively different from those produced in SM di-Higgs production~\cite{Curtin:2013fra,
Ellwanger:2014hia}. To our knowledge, these exotic topologies are not covered by existing ATLAS and CMS di-Higgs searches~\cite{ATLAS:2024combination,CMS:2022dwd,ATLAS:2023bbtt}, and their systematic study is essential for a complete characterization of the di-Higgs signal space at the HL-LHC.

In a companion paper~\cite{mtelba2026}, we performed a scan of $10^6$ semi-constrained NMSSM (scNMSSM) parameter points at large-$\lambda$ and low-$\tan\beta$, identifying 
66 SM-like candidate points with $m_{h_1} \in [122.0, 125.0]$~GeV. 
A key finding of that work is that the Higgs self-coupling modifier $\kappa_\lambda \equiv \lambda_{h_1h_1h_1}/\lambda_{hhh}^{\rm SM}$ is suppressed below unity, spanning $\kappa_\lambda \in [0.884, 0.982]$ with mean $\langle\kappa_\lambda\rangle = 0.944 \pm 0.018$, while the non-resonant di-Higgs cross section lies in the range $\sigma \in [30.3, 36.9]$~fb at $\sqrt{s} = 13.6$~TeV.

This paper is the second in a two-step research programme. In Ref.~\cite{mtelba2026}, we explored the scNMSSM parameter space at large-$\lambda$/low-$\tan\beta$, computed the Higgs self-coupling modifier $\kappa_\lambda$, and determined the non-resonant di-Higgs 
cross section across the 66 candidate SM-like parameter points. The present work builds directly on those results, selecting four representative benchmark points from that parameter space and performing a complete detector-level phenomenological study of the resulting di-Higgs final states at the HL-LHC. The four benchmarks are chosen to be representative of qualitatively distinct phenomenological regimes identified in the scan of Ref.~\cite{mtelba2026}, rather than being selected solely for 
appealing signatures.

We therefore perform a detector-level phenomenological study of the resulting 
di-Higgs final states using 
\textsc{MadGraph5\_aMC@NLO}~\cite{Alwall:2014hca}, 
\textsc{Pythia}~8~\cite{Bierlich:2022pfr}, 
and \textsc{Delphes}~3~\cite{deFavereau:2013fsa}, 
focusing on three qualitatively distinct regimes: exotic multi-$b$, boosted multi-$\tau$, and SM-like di-Higgs production.

It should be emphasised from the outset that the benchmark points considered in this work are not experimentally allowed. As discussed in Section~\ref{subsec:lhc_check}, all four points are in conflict with current LHC constraints on exotic Higgs decays or on the Higgs coupling pattern. The purpose of this study is therefore not to predict the discovery reach of the HL-LHC for a viable scNMSSM scenario, but to characterise the detector-level structure of the exotic di-Higgs topologies themselves. The signal yields quoted below correspond to the stated branching fractions and should not be interpreted as forecasts for experimentally allowed parameter points.

The paper is organized as follows. Section~\ref{sec:model} describes the scNMSSM theoretical framework and the benchmark point selection. Section~\ref{sec:simulation} details the Monte Carlo simulation chain. Section~\ref{sec:analysis} presents 
the signal efficiencies, cut-flow tables, and the characterization of the boosted $a_1$ topology. Section~\ref{sec:background} presents illustrative background estimates and the background suppression that would be required. Section~\ref{sec:discussion} interprets the results and discusses their implications for HL-LHC searches. Section~\ref{sec:conclusion} presents 
our conclusions. 

\section{The scNMSSM and Benchmark Points}
\label{sec:model}

\subsection{Model Overview}
\label{subsec:model}

The semi-constrained Next-to-Minimal Supersymmetric Standard 
Model (scNMSSM) extends the MSSM Higgs sector by a gauge-singlet 
superfield $\hat{S}$, whose vacuum expectation value (VEV) 
generates an effective $\mu$-term $\mu_{\rm eff} = \lambda \langle S \rangle$, 
thereby addressing the $\mu$-problem of the MSSM~\cite{Ellwanger:2009dp}. 
The superpotential of the scNMSSM takes the form:
\begin{equation}
W_{\rm NMSSM} \supset \lambda \hat{S} \hat{H}_u \hat{H}_d 
+ \frac{\kappa}{3} \hat{S}^3,
\end{equation}
where $\lambda$ and $\kappa$ are dimensionless couplings. 
The physical Higgs spectrum consists of three CP-even states 
($h_1, h_2, h_3$), two CP-odd states ($a_1, a_2$), and a 
charged pair ($H^\pm$). In the semi-constrained variant, 
GUT-scale universality is imposed on the soft breaking 
parameters $M_0$, $M_{1/2}$, and $A_0$, while the singlet 
sector parameters $\lambda$, $\kappa$, $A_\lambda$, and 
$A_\kappa$ are treated as free parameters at the electroweak 
scale~\cite{Binjonaid:2014oga}.

In the large-$\lambda$/low-$\tan\beta$ regime considered 
in this work, the lightest CP-even state $h_1$ is identified 
with the observed 125 GeV Higgs boson, and the lightest 
CP-odd state $a_1$ can be sufficiently light to open the 
exotic decay channel $h_1 \to a_1 a_1$. This channel, when 
kinematically accessible, dominates the $h_1$ width and 
gives rise to multi-particle final states that are 
qualitatively distinct from SM di-Higgs 
production~\cite{Curtin:2013fra}.

The trilinear Higgs self-coupling $\lambda_{h_1 h_1 h_1}$ 
receives contributions from both the doublet and singlet 
components of $h_1$, and is conveniently parameterized by 
the modifier:
\begin{equation}
\kappa_\lambda \equiv 
\frac{\lambda_{h_1 h_1 h_1}}{\lambda_{hhh}^{\rm SM}},
\end{equation}
where $\lambda_{hhh}^{\rm SM} = 3m_h^2/v$ with $v = 246$~GeV. 
In our previous work~\cite{mtelba2026}, we demonstrated that $\kappa_\lambda$ is suppressed below unity across the candidate parameter space of this regime, with values spanning $\kappa_\lambda \in [0.884, 0.982]$ and mean 
$\langle \kappa_\lambda \rangle = 0.944 \pm 0.018$.

\subsection{Parameter Scan and Point Selection}
\label{subsec:scan}

The four benchmark points studied in this work are 
selected from a dedicated parameter scan described 
in Ref.~\cite{mtelba2026}. In that scan, $10^6$ 
points in the scNMSSM parameter space were evaluated using \texttt{NMSSMTools}~6.1.2~\cite{Ellwanger:2006rn,
Das:2011dg}, requiring simultaneous satisfaction of:
theoretical consistency (no tachyons, Landau poles, 
or unphysical minima); a Higgs mass window 
$m_{h_1} \in [122, 128]$~GeV, reflecting the 
$\pm 3$~GeV uncertainty in the two-loop mass 
calculation; SM-like Higgs couplings, enforced via 
the singlet fraction $S_{13}^2 < 0.05$ in accordance with LHC signal-strength 
measurements~\cite{ATLAS:2022vkf,CMS:2022dwd}; 
flavor, dark matter, and electroweak precision 
constraints as implemented in 
\texttt{NMSSMTools}~6.1.2.

Of the $10^6$ scanned points, 66 satisfy these requirements, with Higgs masses falling within the tighter window $m_{h_1} \in [122.0, 125.0]$~GeV. The LHC Higgs-sector constraints implemented in \texttt{NMSSMTools} (option \texttt{MODSEL}~19), which include the limits on undetected and exotic Higgs decays and on the Higgs couplings in the $\kappa$ framework, were \emph{not} applied in this scan. We return to this point in Section~\ref{subsec:lhc_check}. From these 66 candidate points, we select four benchmark points (BPs) chosen to span the range of qualitatively distinct di-Higgs final-state topologies accessible in this regime.
\subsection{Benchmark Points}
\label{subsec:benchmarks}

The four benchmark points are selected to span the 
full range of final-state topologies accessible in 
the scNMSSM at large-$\lambda$/low-$\tan\beta$:

\begin{itemize}

\item \textbf{BP1} ($m_{a_1} = 47.4$~GeV): 
The pseudoscalar mass lies above the $b\bar{b}$ 
threshold ($m_{a_1} > 2m_b$), so $a_1 \to b\bar{b}$ 
dominates with $\text{BR}(a_1 \to b\bar{b}) = 91.4\%$. 
The decay chain $h_1 h_1 \to 4a_1 \to 8b$ produces 
a highly boosted multi-$b$ final state.

\item \textbf{BP2} ($m_{a_1} = 8.2$~GeV): 
The pseudoscalar mass lies below the $b\bar{b}$ 
threshold ($m_{a_1} < 2m_b$), so the dominant decay 
is $a_1 \to \tau^+\tau^-$ with 
$\text{BR}(a_1 \to \tau^+\tau^-) = 85.1\%$. 
The resulting topology $h_1 h_1 \to 4a_1 \to 8\tau$ 
features highly collimated $\tau$ pairs 
($\Delta R(\tau\tau) \approx 0.27$) due to the large 
Lorentz boost of the light $a_1$, constituting a distinctive boosted di-tau signature.

\item \textbf{BP3} ($m_{a_1} = 51.9$~GeV): 
With $m_{a_1}$ again above the $b\bar{b}$ threshold, 
this point shares the same dominant decay chain as 
BP1, with $\text{BR}(a_1 \to b\bar{b}) = 91.3\%$. 
The proximity in $m_{a_1}$ between BP1 and BP3 
(47.4 vs.\ 51.9~GeV) allows us to assess the 
stability of the 8$b$ signal efficiency across 
nearby pseudoscalar masses, providing an important 
internal consistency check of our simulation.

\item \textbf{BP4} ($m_{a_1} = 228.6$~GeV): 
The $h_1 \to a_1 a_1$ channel is kinematically 
closed, and the $a_1$ decays predominantly to 
neutralinos ($\text{BR}(a_1 \to \tilde{\chi}_1^0 
\tilde{\chi}_1^0) = 99.3\%$, invisible). 
The $h_1$ decay proceeds entirely through SM-like 
modes ($h_1 \to b\bar{b}$: 71.0\%, 
$h_1 \to WW^*$: 12.1\%, 
$h_1 \to \tau^+\tau^-$: 7.6\%), 
making this benchmark a direct analog of the 
SM HH signal and providing a reference baseline 
for comparison with existing ATLAS and CMS 
di-Higgs searches.

\end{itemize}

The key parameters of all four benchmark points are 
summarized in Table~\ref{tab:benchmarks}, together 
with the di-Higgs production cross section at 
$\sqrt{s} = 13.6$~TeV computed using the 
Carvalho~et~al.\ parametrization~\cite{Carvalho:2015ttv} 
as described in Ref.~\protect\cite{mtelba2026}.

\begin{table}[h!]
\centering
\renewcommand{\arraystretch}{1.3}
\caption{Parameters of the four scNMSSM benchmark points 
selected for di-Higgs phenomenological study at the 
HL-LHC. Cross sections are computed at 
$\sqrt{s} = 13.6$~TeV using the Carvalho 
parametrization~\cite{Carvalho:2015ttv} as described 
in Ref.~\cite{mtelba2026}. Dashes indicate kinematically 
forbidden or negligible channels.}
\label{tab:benchmarks}
\begin{tabular}{lcccc}
\hline\hline
\textbf{Parameter} & \textbf{BP1} & \textbf{BP2} 
                   & \textbf{BP3} & \textbf{BP4} \\
\hline
$m_{h_1}$ [GeV]    
    & 122.1 & 123.3 & 123.2 & 122.3 \\
$m_{a_1}$ [GeV]    
    & 47.4  & 8.2   & 51.9  & 228.6 \\
$\kappa_\lambda$    
    & 0.943 & 0.907 & 0.951 & 0.884 \\
$\sigma(gg\to h_1h_1)$ [fb]            
    & 36.94 & 30.26 & 33.34 & 32.36 \\
\hline
$\text{BR}(h_1\to a_1a_1)$ [\%]       
    & 99.3  & 99.6  & 99.0  & —     \\
$\text{BR}(a_1\to b\bar{b})$ [\%]     
    & 91.4  & —     & 91.3  & —     \\
$\text{BR}(a_1\to\tau^+\tau^-)$ [\%]  
    & 8.1   & 85.1  & 8.2   & —     \\
$\text{BR}(a_1\to gg)$ [\%]           
    & —     & 14.4  & —     & —     \\
$\text{BR}(a_1\to\tilde{\chi}_1^0
  \tilde{\chi}_1^0)$ [\%]
    & —     & —     & —     & 99.3  \\
\hline
$\text{BR}(h_1\to b\bar{b})$ [\%]     
    & —     & —     & —     & 71.0  \\
$\text{BR}(h_1\to WW^*)$ [\%]         
    & —     & —     & —     & 12.1  \\
$\text{BR}(h_1\to\tau^+\tau^-)$ [\%]  
    & —     & —     & —     & 7.6   \\
\hline
Dominant final state                   
    & $8b$  & $8\tau$ & $8b$ & SM-like \\
Role 
    & Exotic $8b$ & Boosted $8\tau$ 
    & Cross-check  & SM reference \\
\hline\hline
\end{tabular}
\end{table}

\subsection{Compatibility with LHC Higgs Constraints (added in v2)}
\label{subsec:lhc_check}

After the benchmark selection, the four points were re-evaluated with \texttt{NMSSMTools}~6.1.2 with the LHC Higgs-constraint option switched on (\texttt{MODSEL}~19~=~1). The results are summarised in Table~\ref{tab:lhccheck}. All four benchmark points are flagged. For BP1--BP3 the undetected branching fraction $\mathrm{BR}(h_1\to\text{undetected})$, which equals $\mathrm{BR}(h_1\to a_1a_1)$ in the decay table of these points, lies between 99.0\% and 99.6\%, and the tool flags it as more than $2\sigma$ away from the fitted value. These points are therefore in tension with the experimental bounds on undetected Higgs decays, whose 95\% CL upper limits are quoted as 12--25\%~\cite{CMS:PAS-NPS-25-005}; see also Refs.~\cite{ATLAS:HiggsMap2022,ATLAS:HiggsComb2026}. BP2 and BP3 are, in addition, flagged as excluded by the combined ATLAS and CMS searches for $h\to aa\to\mu\mu/\tau\tau/bb$.

\begin{table}[H]
\centering
\caption{Result of the \texttt{NMSSMTools} check with the LHC Higgs constraints (\texttt{MODSEL}~19~=~1) for the four benchmark points. $\mathrm{BR}_{\rm und}$ is the branching fraction of $h_1$ into undetected final states, taken from the decay table. Only the flags related to the Higgs sector are listed.}
\label{tab:lhccheck}
\small
\begin{tabular}{@{}llrrp{6.3cm}@{}}
\toprule
BP & scan point & $m_{a_1}$ [GeV] & $\mathrm{BR}_{\rm und}$ [\%] & Flags reported by the tool \\
\midrule
BP1 & np\_0036 & 47.4  & 99.26 & no Higgs boson in the 122.2--128.2~GeV window; $\mathrm{BR}_{\rm und}$ more than $2\sigma$ away \\
BP2 & np\_0078 & 8.2   & 99.57 & excluded by $h\to aa\to\mu\mu/\tau\tau/bb$ (ATLAS+CMS); $\kappa_b$, $\kappa_\tau$, $\mathrm{BR}_{\rm und}$ more than $2\sigma$ away \\
BP3 & np\_0041 & 51.9  & 99.03 & excluded by $h\to aa\to\mu\mu/\tau\tau/bb$ (ATLAS+CMS); $\kappa_b$, $\kappa_\tau$, $\mathrm{BR}_{\rm und}$ more than $2\sigma$ away \\
BP4 & np\_0052 & 228.6 & 0     & $\kappa_b$, $\kappa_\tau$ more than $2\sigma$ away \\
\bottomrule
\end{tabular}
\end{table}

The direct searches give a more specific comparison. The CMS search for $h\to aa$ in the $4\tau$ and $2\mu2\tau$ final states, for $4<m_a<15$~GeV, sets observed 95\% CL upper limits on $\sigma\,\mathrm{BR}(h\to aa\to4\tau)/\sigma_{\rm SM}$ between 0.007 and 0.079 across that mass range~\cite{CMS:2025hh4tau}. For BP2, $\mathrm{BR}(h_1\to a_1a_1)\,\mathrm{BR}(a_1\to\tau^+\tau^-)^2\approx0.9957\times0.851^2\approx0.72$, with SM-like production, which exceeds the weakest limit of the range by about an order of magnitude. For BP1 and BP3, $\mathrm{BR}(h_1\to a_1a_1\to4b)\approx0.83$, whereas the CMS search for $h\to aa\to4b$ with $11<m_a<62.5$~GeV reports observed 95\% CL limits on $\mathrm{BR}(h\to aa\to4b)$ between 0.13\% and 9.4\%, depending on $m_a$~\cite{CMS:PAS-NPS-25-005}. For completeness, the ATLAS search for $h\to aa\to4\tau$ with $15<m_a<60$~GeV sets observed 95\% CL limits on $\mathrm{BR}(h\to aa\to4\tau)$ between 0.06 and 0.23~\cite{ATLAS:2026hh4tau}. BP1 and BP3 lie in this mass range, but with $\mathrm{BR}(a_1\to\tau^+\tau^-)\approx0.08$ their $4\tau$ rate, about $0.65\%$, is well below these limits, so the $4b$ search is the more restrictive one for them; BP2 lies below the ATLAS mass range. These benchmark points therefore lie well above the experimental limits.

Two consequences follow. First, the signal yields presented in the following sections hold for the stated branching fractions only; they do not describe an experimentally allowed point of the scNMSSM. Since the rate for $h_1h_1\to4a_1$ scales as $\mathrm{BR}(h_1\to a_1a_1)^2$, a branching fraction of $10\%$ would reduce all yields by a factor of about $100$, and one of $15\%$ by a factor of about $40$. 
Second, a limited follow-up exploration did not find a replacement. Among 538 points of the original scan with $120<m_{h_1}<130$~GeV, 97 have the decay $h_1\to a_1a_1$ kinematically open, and all of them have $\mathrm{BR}_{\rm und}\geq95\%$. A targeted search over the nine free parameters (7272 model evaluations, seeded from these points) found no point with $122.2<m_{h_1}<128.2$~GeV, $15<m_{a_1}<55$~GeV and $\mathrm{BR}_{\rm und}$ below 98\%. These searches are limited in scope and do not exclude finely tuned regions of parameter space with a smaller $h_1a_1a_1$ coupling. The results below should therefore be read as a characterisation of the detector-level topologies, not as predictions for an allowed scNMSSM point.

\section{Signal Simulation}
\label{sec:simulation}

The signal events for all four benchmark points are generated 
using a full Monte Carlo simulation chain consisting of three 
stages: matrix-element generation, parton showering and 
hadronization, and fast detector simulation.

\subsection{Event Generation}
\label{subsec:generation}

Parton-level signal events for the process 
$gg \to h_1 h_1$ are generated with 
\textsc{MadGraph5\_aMC@NLO} v2.9.21~\cite{Alwall:2014hca} 
at leading order (LO) in QCD, interfaced with the 
\texttt{NNPDF2.3LO} parton distribution 
functions~\cite{Ball:2012cx}. The centre-of-mass 
energy is set to $\sqrt{s} = 13.6$~TeV, consistent 
with LHC Run~3 and HL-LHC conditions. A total of 
$10^5$ unweighted events are generated for each 
benchmark point, with parton shower and hadronization 
disabled at this stage; the resulting Les Houches 
Event (LHE) files serve as input to the subsequent 
showering step.

The $gg \to h_1 h_1$ matrix element is computed 
within the Higgs Effective Field Theory (HEFT) 
framework. The signal cross section for each 
benchmark point is normalized using the Carvalho 
parametrization~\cite{Carvalho:2015ttv} as described 
in Ref.~\cite{mtelba2026}, with the relevant Higgs 
sector parameters extracted directly from the \texttt{NMSSMTools}~6.1.2 spectrum files (Table~\ref{tab:benchmarks}). These parameters are the masses, widths and branching ratios.

\subsection{Decay Injection and Parton Showering}
\label{subsec:shower}

Since \textsc{MadGraph5\_aMC@NLO} does not natively 
propagate the exotic decay chain 
$h_1 \to a_1 a_1 \to \text{multi-parton final states}$ 
through the SLHA decay tables in an automated fashion 
for non-SM particles, we inject the complete 
SLHA-formatted decay tables for $h_1$ and $a_1$ 
directly into the LHE banner prior to showering. 
The injected branching ratios are taken from the 
\texttt{NMSSMTools}~6.1.2 spectrum files and are 
summarized in Table~\ref{tab:benchmarks}.

Parton showering, hadronization, and all subsequent 
$a_1$ decays are handled by 
\textsc{Pythia}~8.316~\cite{Bierlich:2022pfr} 
via the \textsc{DelphesPythia8} interface. 
The $a_1$ mass is set explicitly in the 
\textsc{Pythia} command file through the 
\texttt{SLHA} interface, with mass window 
parameters $[m_{a_1}^{\rm min}, m_{a_1}^{\rm max}]$ 
chosen to reproduce the \texttt{NMSSMTools} 
spectrum for each benchmark point:

\begin{table}[h]
\centering
\caption{Pythia8 mass window parameters used for 
the $a_1$ in each benchmark point.}
\label{tab:pythia}
\renewcommand{\arraystretch}{1.2}
\begin{tabular}{lcccc}
\hline\hline
Parameter [GeV] & BP1 & BP2 & BP3 & BP4 \\
\hline
$m_{a_1}$             & 47.4 & 8.2  & 51.9 & —   \\
$m_{a_1}^{\rm min}$   & 40.0 & 7.0  & 45.0 & —   \\
$m_{a_1}^{\rm max}$   & 55.0 & 9.5  & 59.0 & —   \\
\hline\hline
\end{tabular}
\end{table}

For BP4, no $a_1$ mass window is required since 
$h_1 \to a_1 a_1$ is kinematically closed and 
$h_1$ decays entirely through SM-like modes. 
The \texttt{Main:numberOfEvents} parameter is 
set to $10^5$ in all cases to match the LHE 
event count exactly. All other \textsc{Pythia} 
parameters are kept at their default values, 
including the underlying-event tune 
\texttt{Monash~2013}.
\subsection{Detector Simulation}
\label{subsec:detector}

Fast detector simulation is performed with 
\textsc{Delphes}~3~\cite{deFavereau:2013fsa} 
using the CMS detector response card as 
distributed with \textsc{Delphes}~3, which 
parametrizes the CMS detector geometry and 
performance. The \textsc{DelphesPythia8} 
interface is used to pipe \textsc{Pythia}~8 
output directly into \textsc{Delphes}, 
avoiding intermediate HepMC files and 
reducing disk usage. The total disk usage 
of the simulation output amounts to 
approximately 36~GB across all four 
benchmark points.

Jets are reconstructed using the anti-$k_t$ 
algorithm~\cite{Cacciari:2011ma} with cone 
radius $R = 0.4$. The $b$-tagging working 
point corresponds to an efficiency of 
approximately 70\% for genuine $b$-jets, 
with a light-jet mistag rate of $\sim$1\%, 
consistent with the CMS medium working point. 
Tau-jet identification is performed via the 
\textsc{Delphes} \texttt{TauTagging} module, 
with a reconstruction efficiency of 
$\sim$60\% for hadronically decaying taus.

The simulation produces ROOT output files 
containing reconstructed jets, leptons, 
missing transverse energy ($E_T^{\rm miss}$), 
and the scalar transverse energy sum ($H_T$). 
A total of $4 \times 10^5$ signal events 
($10^5$ per benchmark point) are produced. 
The simulation parameters are summarized in 
Table~\ref{tab:simulation}.

\begin{table}[h]
\centering
\caption{Summary of the Monte Carlo simulation 
chain applied to each benchmark point.}
\label{tab:simulation}
\renewcommand{\arraystretch}{1.2}
\begin{tabular}{lcccc}
\hline\hline
Parameter & BP1 & BP2 & BP3 & BP4 \\
\hline
Generator    
    & \multicolumn{4}{c}
      {\textsc{MadGraph5\_aMC@NLO} v2.9.21} \\
Shower/Decay 
    & \multicolumn{4}{c}
      {\textsc{Pythia}~8.316} \\
Detector sim.
    & \multicolumn{4}{c}
      {\textsc{Delphes}~3 (CMS card)} \\
$\sqrt{s}$ [TeV]     
    & \multicolumn{4}{c}{13.6} \\
Events generated     
    & \multicolumn{4}{c}{$10^5$} \\
\hline
$\sigma(gg\to h_1h_1)$ [fb]
    & 36.94 & 30.26 & 33.34 & 32.36 \\
Dominant topology    
    & $8b$ & $8\tau$ & $8b$ & SM-like \\
\hline\hline
\end{tabular}
\end{table}

The analysis of the ROOT output is performed 
using \texttt{uproot}~\cite{Pivarski:2021} 
and \texttt{NumPy}~\cite{numpy}, with object 
selection criteria described in 
Section~\ref{sec:analysis}.

\section{Signal Characterization}
\label{sec:analysis}

\subsection{Object Selection}
\label{subsec:selection}

Reconstructed objects are selected from the 
\textsc{Delphes} output according to the 
following criteria, chosen to reflect realistic 
HL-LHC analysis conditions while maintaining 
high signal acceptance across all four benchmark 
points.

\textbf{Jets} are required to satisfy:
\begin{itemize}
\item Transverse momentum $p_T > 25$~GeV
\item Pseudorapidity $|\eta| < 2.5$
\end{itemize}

\textbf{$b$-tagged jets} are identified using 
the \textsc{Delphes} \texttt{BTagging} module 
with the CMS medium working point, corresponding 
to an efficiency of $\sim$70\% for genuine 
$b$-jets. A jet is classified as $b$-tagged if 
its \texttt{BTag} flag satisfies 
$\texttt{BTag} \geq 1$. We note that the 
$b$-tagging efficiency uncertainty of $\sim$5\% 
per jet propagates to a signal efficiency 
uncertainty of $\sim$20\% for the $\geq 4b$ 
selection, which is accounted for within the 
systematic uncertainty $\delta = 20\%$ adopted 
in Section~\ref{sec:background}.

\textbf{$\tau$-tagged jets} are identified 
using the \textsc{Delphes} \texttt{TauTagging} 
module. A jet is classified as a $\tau$-jet if 
its \texttt{TauTag} flag satisfies 
$\texttt{TauTag} \geq 1$, with:
\begin{itemize}
\item Transverse momentum $p_T > 20$~GeV
\item Pseudorapidity $|\eta| < 2.5$
\end{itemize}

The lower $p_T$ threshold for $\tau$-jets 
reflects the softer momentum spectrum expected 
from the boosted $a_1 \to \tau^+\tau^-$ decay, 
particularly for BP2 where $m_{a_1} = 8.2$~GeV.

\textbf{Scalar transverse energy} $H_T$ is 
computed from the \textsc{Delphes} 
\texttt{ScalarHT} module as the scalar sum of 
all reconstructed jet transverse momenta.

No explicit lepton veto is applied, as the 
signal topologies under consideration are 
dominated by hadronic final states.

\subsection{Signal Efficiencies}
\label{subsec:efficiencies}

Signal efficiencies are defined as the fraction 
of simulated events satisfying a given 
selection requirement:
\begin{equation}
\varepsilon = \frac{N_{\rm sel}}{N_{\rm total}},
\end{equation}
where $N_{\rm total} = 10^5$ for each benchmark 
point and $N_{\rm sel}$ is the number of events 
passing the selection. The expected signal yield 
at the HL-LHC with an integrated luminosity of 
$\mathcal{L} = 3000$~fb$^{-1}$ is:
\begin{equation}
S = \sigma(gg \to h_1 h_1) \times 
    \mathcal{L} \times \varepsilon.
\end{equation}

Table~\ref{tab:efficiencies} summarizes the 
signal efficiencies and expected event yields 
for the primary selection criteria applied 
to each benchmark point.

\begin{table}[h!]
\centering
\renewcommand{\arraystretch}{1.3}
\caption{Signal efficiencies $\varepsilon$ and 
expected event yields $S$ at 
$\mathcal{L} = 3000$~fb$^{-1}$ for the primary 
selection criteria applied to each benchmark 
point. Object selection criteria are described 
in Section~\ref{subsec:selection}.}
\label{tab:efficiencies}
\begin{tabular}{llccc}
\hline\hline
BP & Selection & $\varepsilon$ [\%] & 
     $S$ & Physics channel \\
\hline
BP1 & $\geq 2b$ & 70.6 & 78,195 & 
      $h_1h_1\to 4a_1\to 8b$ \\
BP1 & $\geq 4b$ & 14.3 & 15,814 & 
      $h_1h_1\to 4a_1\to 8b$ \\
\hline
BP2 & $\geq 1\tau$ & 78.5 & 71,245 & 
      $h_1h_1\to 4a_1\to 8\tau$ \\
BP2 & $\geq 2\tau$ & 38.4 & 34,814 & 
      $h_1h_1\to 4a_1\to 8\tau$ \\
BP2 & $\geq 3\tau$ & 10.0 &  9,087 & 
      $h_1h_1\to 4a_1\to 8\tau$ \\
\hline
BP3 & $\geq 2b$ & 70.2 & 70,219 & 
      $h_1h_1\to 4a_1\to 8b$ \\
BP3 & $\geq 4b$ & 14.0 & 13,998 & 
      $h_1h_1\to 4a_1\to 8b$ \\
\hline
BP4 & $\geq 2b$         & 50.6 & 49,093 & 
      SM-like \\
BP4 & $\geq 2b+\geq 1\tau$ &  7.7 &  7,504 & 
      SM-like ($b\bar{b}\tau\tau$) \\
\hline\hline
\end{tabular}
\end{table}

Several observations are noteworthy. First, 
the exotic 8$b$ benchmarks BP1 and BP3 yield 
comparable $b$-jet efficiencies of 70.6\% and 
70.2\% respectively at the $\geq 2b$ level, 
reflecting their similar $a_1$ masses 
(47.4 vs.\ 51.9~GeV) and identical dominant 
decay mode. This internal consistency provides 
a cross-check of the simulation pipeline. 
Second, the SM-like benchmark BP4 yields a 
$\geq 2b$ efficiency of 50.6\%, consistent 
with expectations from SM HH$\to b\bar{b}b\bar{b}$ 
studies in the literature~\cite{ATLAS:2023bbbb,
CMS:2022dwd}. Third, the boosted benchmark BP2 
shows a $\geq 2\tau$ efficiency of 38.4\%, 
representing a factor of $\sim$21 enhancement 
over the SM HH baseline of $\sim$1.8\%, a 
consequence of the highly collimated 
$a_1\to\tau^+\tau^-$ topology discussed in 
Section~\ref{subsec:boosted}.

\subsection{Cut-Flow Analysis}
\label{subsec:cutflow}

Table~\ref{tab:cutflow} presents the full 
cut-flow for each benchmark point, showing 
the progressive effect of each selection 
requirement on the signal event count and 
efficiency.

\begin{table*}[h!]
\centering
\renewcommand{\arraystretch}{1.2}
\caption{Cut-flow for all four benchmark points 
at $\mathcal{L} = 3000$~fb$^{-1}$. 
$\varepsilon$ is the signal efficiency and 
$S$ is the expected yield.}
\label{tab:cutflow}
\resizebox{\textwidth}{!}{%
\begin{tabular}{lcccccccc}
\hline\hline
& \multicolumn{2}{c}{\textbf{BP1} ($8b$)} 
& \multicolumn{2}{c}{\textbf{BP2} ($8\tau$)}
& \multicolumn{2}{c}{\textbf{BP3} ($8b$)}
& \multicolumn{2}{c}{\textbf{BP4} (SM)} \\
\cmidrule(lr){2-3}\cmidrule(lr){4-5}
\cmidrule(lr){6-7}\cmidrule(lr){8-9}
\textbf{Selection} & 
$\varepsilon$[\%] & $S$ &
$\varepsilon$[\%] & $S$ &
$\varepsilon$[\%] & $S$ &
$\varepsilon$[\%] & $S$ \\
\hline
All events    
& 100.0 & 110,820 
& 100.0 &  90,780 
& 100.0 & 100,020 
& 100.0 &  97,080 \\
$\geq 2$ jets    
&  97.6 & 108,176 
&  92.5 &  83,930 
&  97.5 &  97,529 
&  96.8 &  93,991 \\
\hline
$\geq 2b$     
&  70.6 &  78,190 
&  — & — 
&  70.2 &  70,219 
&  50.6 &  49,093 \\
$\geq 2b$+$H_T\!>\!150$
&  66.5 &  73,693 
&  — & — 
&  66.0 &  66,034 
&  49.3 &  47,902 \\
$\geq 3b$     
&  38.7 &  42,877 
&  — & — 
&  38.4 &  38,417 
&  — & — \\
$\geq 4b$     
&  14.3 &  15,814 
&  — & — 
&  14.0 &  13,998 
&  — & — \\
$\geq 4b$+$H_T\!>\!200$  
&  13.5 &  14,964 
&  — & — 
&  13.2 &  13,219 
&  — & — \\
\hline
$\geq 1\tau$  
&  — & — 
&  78.5 &  71,245 
&  — & — 
&  — & — \\
$\geq 2\tau$  
&  — & — 
&  38.4 &  34,814 
&  — & — 
&  — & — \\
$\geq 2\tau$+$H_T\!>\!100$ 
&  — & — 
&  37.6 &  34,089 
&  — & — 
&  — & — \\
$\geq 3\tau$  
&  — & — 
&  10.0 &   9,087 
&  — & — 
&  — & — \\
$\geq 3\tau$+$H_T\!>\!100$ 
&  — & — 
&   9.97 &   9,051 
&  — & — 
&  — & — \\
\hline
$\geq 2b$+$\geq 1\tau$ 
&  — & — 
&  — & — 
&  — & — 
&   7.7 &   7,504 \\
$\geq 2b$+$\geq 1\tau$+$H_T\!>\!150$ 
&  — & — 
&  — & — 
&  — & — 
&   7.6 &   7,390 \\
$\geq 2b$+$\geq 2\tau$ 
&  — & — 
&  — & — 
&  — & — 
&   0.87 &     848 \\
\hline\hline
\end{tabular}%
}
\end{table*}

\subsection{Kinematic Distributions}
\label{subsec:kinematics}

To further characterize the signal 
topologies, we examine the key kinematic 
distributions for all four benchmark 
points after the basic jet selection 
($\geq 2$ jets, $p_T > 25$~GeV, 
$|\eta| < 2.5$).

\subsubsection{Scalar Transverse Energy}

\begin{figure}[h!]
\centering
\includegraphics[width=0.65\textwidth]
    {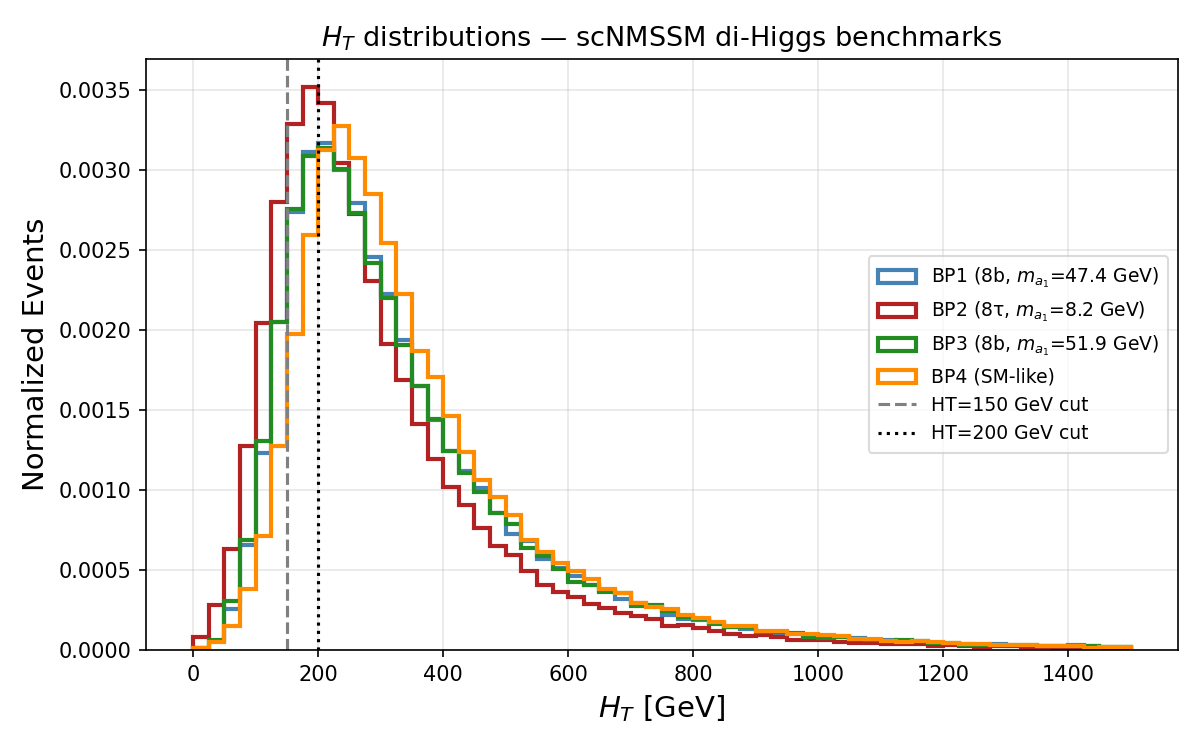}
\caption{Scalar transverse energy 
$H_T$ distributions for all four 
benchmark points after requiring 
$\geq 2$ jets with $p_T > 25$~GeV 
and $|\eta| < 2.5$. The vertical 
dashed lines indicate the 
$H_T > 150$~GeV and $H_T > 200$~GeV 
selection thresholds applied in the 
analysis. All distributions are 
normalized to unity.}
\label{fig:HT}
\end{figure}

Figure~\ref{fig:HT} shows the $H_T$ 
distributions for all four benchmark 
points. The exotic $8b$ benchmarks 
BP1 and BP3 extend to higher $H_T$ 
values than BP4, reflecting the larger 
number of energetic jets per event 
from the $h_1 h_1 \to 4a_1 \to 8b$ 
decay chain. BP2 shows a softer $H_T$ 
spectrum, consistent with the 
collimated and partially merged 
$\tau$-pair topology where individual 
jet energies are lower. The vertical 
lines indicate the $H_T > 150$~GeV 
and $H_T > 200$~GeV thresholds 
applied in the analysis.

\subsubsection{Jet Multiplicity}

\begin{figure*}[h!]
\centering
\includegraphics[width=\textwidth]
    {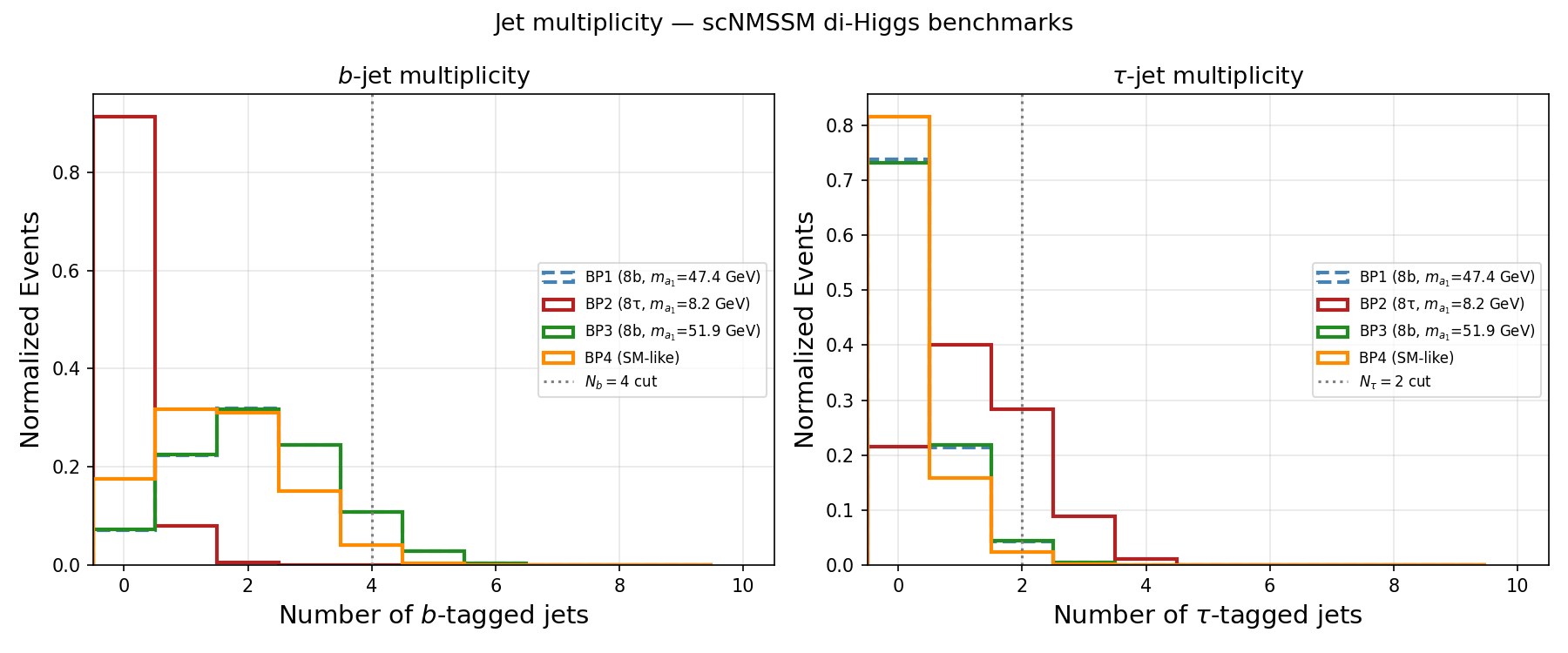}
\caption{Distributions of the number 
of $b$-tagged jets ($N_b$, left) and 
$\tau$-tagged jets ($N_\tau$, right) 
for all four benchmark points after 
basic object selection. The vertical 
dashed lines indicate the 
$N_b = 4$ and $N_\tau = 2$ selection 
thresholds used in the analysis. 
BP1 (blue dashed) and BP3 (green) 
overlap in the $b$-jet distribution, 
reflecting their similar $a_1$ masses. 
All distributions are normalized 
to unity.}
\label{fig:multiplicity}
\end{figure*}

Figure~\ref{fig:multiplicity} shows 
the $b$-tagged jet and $\tau$-tagged 
jet multiplicity distributions for 
all four benchmark points. The 
topological separation between the 
benchmarks is clearly visible: BP1 
and BP3 produce the highest $b$-jet 
multiplicities, with distributions 
extending to $N_b \geq 6$, while 
BP2 dominates in $\tau$-jet 
multiplicity, with events containing 
up to four resolved $\tau$-tagged 
jets. BP4 follows the expected 
SM-like pattern, peaking at 
$N_b = 1$--$2$ with very few 
$\tau$-tagged jets. The close 
agreement between BP1 and BP3 in 
the $b$-jet multiplicity distribution 
(dashed blue vs.\ solid green) 
reflects their similar $a_1$ masses 
and decay kinematics, providing an 
internal consistency check of the 
simulation.

\subsubsection{Transverse Momentum}

\begin{figure*}[h!]
\centering
\includegraphics[width=0.95\textwidth]
    {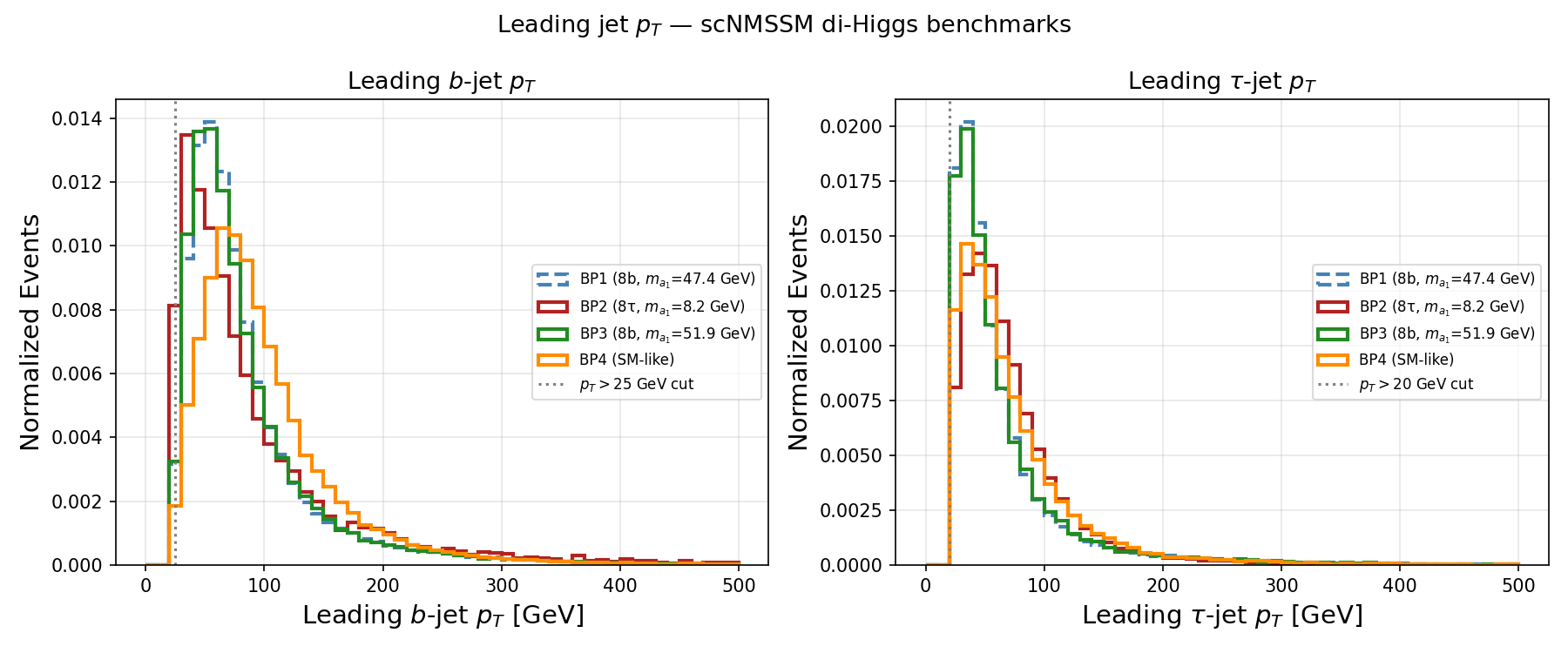}
\caption{Transverse momentum $p_T$ 
distributions of the leading 
$b$-tagged jet (left) and the 
leading $\tau$-tagged jet (right) 
for all four benchmark points 
after basic object selection. 
The vertical dashed lines indicate 
the $p_T > 25$~GeV threshold for 
$b$-jets and the $p_T > 20$~GeV 
threshold for $\tau$-jets. 
All distributions are normalized 
to unity.}
\label{fig:pT}
\end{figure*}

Figure~\ref{fig:pT} shows the 
transverse momentum distributions 
of the leading $b$-tagged and 
$\tau$-tagged jets. For the 
$b$-jet $p_T$ (left panel), 
BP1 and BP3 show harder spectra 
than BP4, consistent with the 
higher $b$-jet multiplicity and 
the boost imparted by the 
$h_1 h_1 \to 4a_1$ decay chain. 
For the $\tau$-jet $p_T$ (right 
panel), BP2 peaks at lower values 
than the other benchmarks, 
reflecting the soft momentum 
spectrum of the boosted and 
partially merged $\tau$ pairs 
from the light $a_1$ with 
$m_{a_1} = 8.2$~GeV. The 
vertical dashed lines indicate 
the $p_T > 25$~GeV and 
$p_T > 20$~GeV thresholds 
applied for $b$-jets and 
$\tau$-jets respectively.

\subsection{Boosted Topology of BP2}
\label{subsec:boosted}

The most distinctive feature of BP2 is the 
highly boosted nature of the $a_1\to\tau^+\tau^-$ 
decay. For $m_{a_1} = 8.2$~GeV, the pseudoscalar 
is produced with a typical Lorentz boost 
$\gamma \approx m_{h_1}/(2m_{a_1}) \approx 7.5$, 
causing the two $\tau$ leptons to be emitted 
with a characteristic angular separation:
\begin{equation}
\Delta R(\tau,\tau) \approx \frac{2}{\gamma} 
\approx 0.27,
\end{equation}
which lies below the standard jet cone radius 
of $R = 0.4$.

\begin{figure}[h!]
\centering
\includegraphics[width=0.65\textwidth]
    {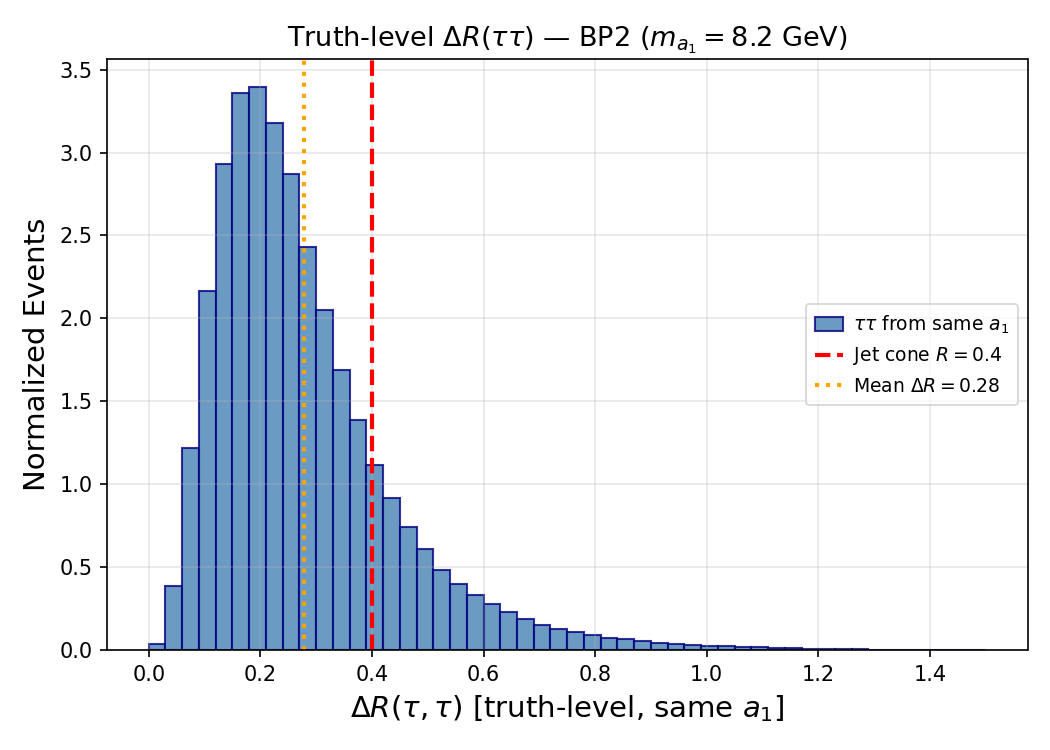}
\caption{Truth-level $\Delta R(\tau,\tau)$ 
distribution for $\tau$ pairs originating 
from the same $a_1$ decay in BP2 
($m_{a_1} = 8.2$~GeV, $10^5$ events). 
The vertical dashed line indicates the 
standard jet cone radius $R = 0.4$. 
A total of 82.4\% of $\tau$ pairs fall 
below this threshold, demonstrating the 
highly boosted nature of the 
$a_1\to\tau^+\tau^-$ decay.}
\label{fig:deltaR}
\end{figure}

To quantify this effect, we compute the 
$\Delta R$ separation between all pairs of 
truth-level $\tau$ leptons originating from 
the same $a_1$ decay across the $10^5$ 
simulated BP2 events. The resulting 
distribution is shown in 
Fig.~\ref{fig:deltaR}. We find:
\begin{itemize}
\item Mean $\Delta R(\tau,\tau) = 0.278$
\item Median $\Delta R(\tau,\tau) = 0.240$
\item $82.4\%$ of same-$a_1$ $\tau$ pairs 
      have $\Delta R < 0.4$
\item $65.9\%$ have $\Delta R < 0.3$
\item $37.1\%$ have $\Delta R < 0.2$
\end{itemize}

Since 82.4\% of $\tau$ pairs from the same 
$a_1$ fall within a single jet cone, standard 
$\tau$-tagging algorithms are designed to reconstruct isolated, resolved $\tau$ jets and are therefore largely ineffective for this topology. 
This is consistent with the observed drop in efficiency from $\varepsilon(\geq 1\tau) = 78.5\%$ 
to $\varepsilon(\geq 2\tau) = 38.4\%$: a 
significant fraction of events contain merged 
$\tau$ pairs that are reconstructed as a 
single $\tau$-tagged jet rather than two 
separate objects.

This boosted topology is, to our knowledge, qualitatively distinct from the final states targeted by existing ATLAS and CMS di-Higgs searches~\cite{ATLAS:2024combination,
CMS:2022dwd,ATLAS:2023bbtt} and motivates 
the development of dedicated boosted 
$a_1$-tagging strategies, as discussed 
in Section~\ref{sec:discussion}.

\section{Illustrative Background Estimates and Required Suppression}
\label{sec:background}

The estimates in this section are intended solely to illustrate the scale of the background rejection that would be required for a discovery claim. They do not constitute a sensitivity projection, since the benchmark points are not experimentally allowed and the background model is simplified.

A full Monte Carlo simulation of the SM background 
processes is beyond the scope of this phenomenological 
study. Instead, we adopt background cross sections 
from published ATLAS and CMS analyses at 
$\sqrt{s} = 13$~TeV, which we use to derive illustrative estimates of the expected background yields and of the background suppression that a discovery would require at the HL-LHC. In view of Section~\ref{subsec:lhc_check}, these estimates do not describe an experimentally allowed scenario.

\subsection{Background Processes}
\label{subsec:backgrounds}

The dominant background processes for each 
signal topology are as follows.

\textbf{For the $8b$ channels (BP1, BP3):}
The primary backgrounds are QCD multijet 
production and $t\bar{t}$ pair production, 
both of which can yield multiple $b$-tagged 
jets. Following Refs.~\cite{ATLAS:2023bbbb,
CMS:2022dwd}, we estimate the total background 
in the $\geq 4b$ signal region to be in the 
range $B \sim 3{,}000$--$6{,}000$ events at 
$\mathcal{L} = 3000$~fb$^{-1}$, after 
applying representative kinematic selections.

\textbf{For the boosted $8\tau$ channel (BP2):}
The dominant backgrounds are 
$Z/\gamma^*\to\tau^+\tau^-$+jets, 
$t\bar{t}$, and $ZZ\to 4\tau$. 
For the $\geq 2\tau$ selection, the total 
background is estimated to be in the range 
$B \sim 40{,}000$--$70{,}000$ events, 
while for the $\geq 3\tau$ selection it 
reduces to $B \sim 5{,}000$--$15{,}000$ 
events, based on cross-section estimates 
from published ATLAS and CMS analyses 
after representative kinematic 
selections~\cite{ATLAS:2023bbtt}.

\textbf{For the SM-like channel (BP4):}
The dominant backgrounds mirror those of 
the SM HH analyses: QCD multijet and 
$t\bar{t}$ for the $\geq 2b$ channel, 
and $t\bar{t}Z$ for the 
$\geq 2b$+$\geq 1\tau$ channel. 
The total background in the $\geq 2b$ 
signal region is estimated at 
$B \sim 60{,}000$--$180{,}000$ events, 
and in the $\geq 2b$+$\geq 1\tau$ region 
at $B \sim 3{,}000$--$8{,}000$ events.

\subsection{Statistical Significance}
\label{subsec:significance}

The expected statistical significance is 
estimated using:
\begin{equation}
Z_{\rm stat} = \frac{S}{\sqrt{B}},
\end{equation}
and the significance including a fractional 
systematic uncertainty $\delta$ on the 
background:
\begin{equation}
Z_{\rm syst} = \frac{S}{\sqrt{B + 
(\delta \cdot B)^2}},
\end{equation}
where we adopt $\delta = 20\%$ as an illustrative effective background uncertainty for the simplified sensitivity estimate. This parameter is not a substitute for a channel-specific 
experimental uncertainty model, which would require dedicated background simulation and data-driven validation. 
The resulting significance should be  interpreted as a sensitivity diagnostic rather than an official HL-LHC projection~\cite{ATLAS:2024combination,
CMS:2022dwd}. The results are summarized 
in Table~\ref{tab:significance}.

\begin{table}[h!]
\centering
\renewcommand{\arraystretch}{1.3}
\caption{Expected signal yields $S$, 
background estimates $B$, and significances 
at $\mathcal{L} = 3000$~fb$^{-1}$. 
$Z_{\rm stat} = S/\sqrt{B}$ and 
$Z_{\rm syst} = S/\sqrt{B+(\delta B)^2}$ 
with $\delta = 20\%$. Background estimates 
are derived from published ATLAS and CMS 
analyses (see text).}
\label{tab:significance}
\resizebox{\columnwidth}{!}{%
\begin{tabular}{llccccc}
\hline\hline
BP & Selection & $S$ & 
    $B_{\rm low}$ & $B_{\rm high}$ & 
    $Z_{\rm stat}$ & $Z_{\rm syst}$ \\
\hline
BP1 & $\geq 4b$       
    & 15,814 & 3,000  & 6,000   
    & 204--289 & 13.2--26.3 \\
BP3 & $\geq 4b$       
    & 13,998 & 3,000  & 6,000   
    & 181--256 & 11.6--23.2 \\
BP2 & $\geq 2\tau$    
    & 34,814 & 40,000 & 70,000 
    & 132--174 & 2.5--4.4   \\
BP2 & $\geq 3\tau$    
    &  9,087 & 5,000  & 15,000  
    & 74--129  & 3.0--9.1   \\
BP4 & $\geq 2b$       
    & 49,093 & 60,000 & 180,000 
    & 116--200 & 1.4--4.1   \\
BP4 & $\geq 2b{+}\geq 1\tau$ 
    &  7,504 & 3,000  & 8,000   
    & 84--137  & 4.7--12.5  \\
\hline\hline
\end{tabular}%
}
\end{table}

\subsection{Required Background Suppression}
\label{subsec:suppression}

The values of $B_{\mathrm{req}}$ derived below quantify the scale of the required background rejection and should not be read as a statement that such a discovery is achievable, given the limitations of the background model and the excluded status of the benchmark points. For the channels where the illustrative significance estimate $Z_{\rm syst}$ falls below the nominal $5\sigma$ discovery threshold, 
we compute the maximum background level 
$B_{\rm req}$ consistent with a $5\sigma$ 
discovery at $\delta = 20\%$:
\begin{equation}
\frac{S}{\sqrt{B_{\rm req} + 
(\delta \cdot B_{\rm req})^2}} = 5,
\end{equation}
which gives the quadratic equation 
$25\delta^2 B_{\rm req}^2 + 
25 B_{\rm req} - S^2 = 0$, 
with positive solution:
\begin{equation}
B_{\rm req} = \frac{-25 + 
\sqrt{625 + 4 \cdot 25\delta^2 \cdot S^2}}
{2 \cdot 25\delta^2}.
\end{equation}

The results are summarized in 
Table~\ref{tab:suppression}.

\begin{table}[h!]
\centering
\renewcommand{\arraystretch}{1.3}
\caption{Required maximum background 
$B_{\rm req}$ for $5\sigma$ discovery 
with $\delta = 20\%$ systematic uncertainty, 
and the corresponding effective cross section 
$\sigma_{\rm eff} = B_{\rm req}/\mathcal{L}$ 
at $\mathcal{L} = 3000$~fb$^{-1}$.}
\label{tab:suppression}
\begin{tabular}{llccc}
\hline\hline
BP & Selection & $S$ & 
    $B_{\rm req}$ & 
    $\sigma_{\rm eff}$ [fb] \\
\hline
BP1 & $\geq 4b$            
    & 15,814 & 15,802 & 5.27 \\
BP3 & $\geq 4b$            
    & 13,998 & 13,986 & 4.66 \\
BP2 & $\geq 2\tau$         
    & 34,814 & 34,802 & 11.60 \\
BP2 & $\geq 3\tau$         
    &  9,087 &  9,075 &  3.02 \\
BP4 & $\geq 2b$            
    & 49,093 & 49,081 & 16.36 \\
BP4 & $\geq 2b{+}\geq 1\tau$ 
    &  7,504 &  7,492 &  2.50 \\
\hline\hline
\end{tabular}
\end{table}

The results reveal a clear hierarchy between 
the exotic and SM-like channels. For BP1 and 
BP3, the required background $B_{\rm req} \approx 14{,}000$--$16{,}000$
events exceeds the literature estimate of 
$B \sim 3{,}000$--$6{,}000$ events in the 
$\geq 4b$ signal region, which is a comparison of orders of magnitude only. It does not establish that a $5\sigma$ discovery is achievable: the background model is simplified, and the benchmark points are excluded by the constraints discussed in Section~\ref{subsec:lhc_check}.

For BP2, the situation is qualitatively 
different. In the $\geq 3\tau$ channel, 
a $5\sigma$ discovery requires the total 
background to not exceed 
$B_{\rm req} \approx 9{,}075$ events 
($\sigma_{\rm eff} < 3.0$~fb). The dominant irreducible backgrounds in the 
$\geq 3\tau$ channel are $Z/\gamma^*\to 
\tau^+\tau^-$+jets and $t\bar{t}$, both of 
which produce genuine $\tau$-rich final states. 
The $ZZ\to 4\tau$ process, with 
$\sigma(ZZ\to 4\tau) = \sigma(ZZ)\times 
[\text{BR}(Z\to\tau^+\tau^-)]^2 \approx 
16~\text{pb}\times(0.034)^2 \approx 18$~fb, 
contributes a smaller but non-negligible 
background of $\mathcal{O}(50)$ events at 
$\mathcal{L}=3000$~fb$^{-1}$ after basic 
kinematic selections. The dominant challenge 
for the $\geq 3\tau$ channel is therefore 
the suppression of $Z/\gamma^*\to\tau^+\tau^-$+jets 
and QCD fake-$\tau$ backgrounds, which 
together drive the total background estimate 
of $B\sim 5{,}000$--$15{,}000$ events 
cited above. This confirms that dedicated 
boosted $a_1$ taggers are required to 
achieve the necessary background rejection. This indicates that, under the present background model, the boosted $8\tau$ topology cannot be established with 
standard $\tau$-tagging alone, and 
motivates the development of dedicated 
boosted $a_1$ taggers exploiting the 
merged $\tau$-pair substructure identified 
in Section~\ref{subsec:boosted}.

For BP4, the $\geq 2b$+$\geq 1\tau$ 
channel yields $B_{\rm req} \approx 7{,}492$ 
events, consistent with the literature 
background estimate of 
$B \sim 3{,}000$--$8{,}000$ events, 
suggesting a marginal illustrative sensitivity, for a point that is itself not experimentally allowed. This is consistent with 
existing HL-LHC projections for SM 
HH production~\cite{Cepeda:2019klc,
ATL-PHYS-PUB-2022-005}, and serves 
as a cross-check of our simulation pipeline.

We emphasize that all significance estimates presented here are indicative rather than definitive, as they rely on background yields derived from published cross-section measurements rather than a dedicated background simulation. 
The assumed systematic uncertainty of $\delta = 20\%$ is conservative relative to current ATLAS and CMS multi-jet analyses~\cite{ATLAS:2024combination,
CMS:2022dwd}, but may underestimate the full systematic budget at the HL-LHC when trigger efficiencies, pile-up effects, and object mis-identification rates are fully accounted for. A complete Monte Carlo background estimation with 
optimized selections and data-driven methods is required for a quantitative sensitivity assessment, and is left for future work.

\subsection{Limitations of the Background Model}
\label{subsec:limitations}

The background and significance estimates 
presented in this section are intended to quantify the scale of the required background rejection and to compare the relative promise of the benchmark channels. The following limitations should be noted:
\begin{itemize}
\item Background yields are estimated from 
      published ATLAS and CMS cross-section 
      measurements rather than a dedicated 
      background Monte Carlo simulation.
\item The event selections used here are 
      not identical to those of the published 
      analyses from which the background 
      estimates are derived.
\item Fake-$\tau$ rates, pileup effects, 
      and trigger efficiencies are not 
      explicitly modelled.
\item The 20\% background uncertainty 
      $\delta$ is an illustrative effective 
      parameter and does not represent a 
      complete systematic uncertainty budget.
\item The resulting significances are not 
      official ATLAS or CMS sensitivity 
      projections.
\end{itemize}
A dedicated analysis using simulated backgrounds, realistic object and trigger efficiencies, pileup treatment, and data-driven background constraints will be required before discovery claims can 
be assessed quantitatively.

In addition to the limitations of the background model, the benchmark points themselves are not experimentally allowed. The significance estimates presented above therefore do not describe the discovery potential of the HL-LHC for a viable scNMSSM scenario. They serve only to compare the relative scale of the required background suppression across the considered topologies under a common, simplified set of assumptions.

\section{Discussion}
\label{sec:discussion}

The results presented in this work 
demonstrate that di-Higgs production in 
the scNMSSM at large-$\lambda$/low-$\tan\beta$ 
gives rise to qualitatively distinct 
multi-particle final states that differ 
substantially from those targeted by 
existing ATLAS and CMS di-Higgs searches. 
We discuss below the key physics implications 
of our findings for each signal topology.

\subsection{Exotic Multi-$b$ Channels 
            (BP1 and BP3)}
\label{subsec:disc_8b}

The decay chain $h_1 h_1 \to 4a_1 \to 8b$ 
produces a final state with up to eight 
$b$-quarks, a topology with no direct analog 
in the SM. The high $b$-jet multiplicity 
arises from the combination of a large 
$\text{BR}(h_1 \to a_1 a_1) \approx 99\%$ 
and a dominant 
$\text{BR}(a_1 \to b\bar{b}) \approx 91\%$ 
for $m_{a_1}$ above the $b\bar{b}$ threshold.

Our simulation yields $b$-jet efficiencies 
of $\varepsilon(\geq 2b) \approx 70\%$ and 
$\varepsilon(\geq 4b) \approx 14\%$ for 
both BP1 and BP3, consistent with each 
other given the proximity of their $a_1$ 
masses (47.4 vs.\ 51.9~GeV). This internal 
consistency provides confidence in the 
robustness of our simulation pipeline.

The $\geq 4b$ efficiency of $\sim$14\% 
reflects the higher parton-level $b$-quark 
multiplicity (up to 8 per event) relative 
to the SM HH$\to b\bar{b}b\bar{b}$ process 
(4 per event). Under conservative background 
estimates from the 
literature~\cite{ATLAS:2023bbbb,CMS:2022dwd}, 
the required background suppression for 
$5\sigma$ discovery 
($B_{\rm req} \approx 14{,}000$--$16{,}000$ 
events) exceeds the estimated background of 
$B \sim 3{,}000$--$6{,}000$ events in the 
$\geq 4b$ signal region, which is an illustrative comparison only. Because the benchmark points are in conflict with the limits on exotic Higgs decays (Section~\ref{subsec:lhc_check}), it does not support a statement on the discovery reach of the HL-LHC.

\subsection{Boosted Multi-$\tau$ Channel 
            (BP2)}
\label{subsec:disc_8tau}

The most phenomenologically novel result 
of this work is the boosted $8\tau$ 
topology of BP2. With $m_{a_1} = 8.2$~GeV, 
the pseudoscalar is produced with a typical 
Lorentz boost 
$\gamma \approx m_{h_1}/(2m_{a_1}) \approx 7.5$, 
causing the $\tau$ pairs from each 
$a_1 \to \tau^+\tau^-$ decay to be highly 
collimated with a characteristic angular 
separation $\Delta R(\tau,\tau) \approx 0.27$. 
As demonstrated in 
Section~\ref{subsec:boosted}, 82.4\% of 
same-$a_1$ $\tau$ pairs have $\Delta R < 0.4$ 
and are therefore merged within a single 
reconstructed jet cone of radius $R = 0.4$.

This boosted topology has two important 
consequences. First, the $\geq 2\tau$ 
efficiency of 38.4\% represents a factor 
of $\sim$21 enhancement over the SM HH 
baseline of $\sim$1.8\%~\cite{mtelba2026}, 
arising from the dramatically increased 
$\tau$-jet rate per event when 
$\text{BR}(a_1 \to \tau^+\tau^-) \approx 
85\%$ with four $a_1$ per event. 
Second, the merging of $\tau$ pairs within 
single jet cones means that the effective 
number of resolved $\tau$-tagged jets is 
significantly reduced relative to the 
parton-level expectation of eight $\tau$ 
leptons per event, directly explaining 
the efficiency drop from 
$\varepsilon(\geq 1\tau) = 78.5\%$ to 
$\varepsilon(\geq 2\tau) = 38.4\%$.

This topology is not covered by any 
existing ATLAS or CMS di-Higgs 
search~\cite{ATLAS:2024combination,
CMS:2022dwd,ATLAS:2023bbtt}. Standard 
$\tau$-tagging algorithms are optimized 
for isolated, resolved $\tau$ jets and 
are not designed to identify merged 
$\tau$-pair systems with 
$\Delta R \lesssim 0.4$. The development of dedicated boosted $a_1$ taggers is therefore strongly motivated. These taggers would be analogous to the boosted Higgs and boosted top taggers widely used in LHC analyses. They could exploit 
the characteristic sub-jet structure 
of the merged $\tau$-pair system, 
including the invariant mass of the 
merged system 
($m_{\tau\tau} \approx m_{a_1} = 8.2$~GeV), 
the sub-jet multiplicity, and the 
radiation pattern within the fat jet.

The $\geq 3\tau$ selection requires three 
resolved $\tau$-tagged jets and reduces 
the signal efficiency to 
$\varepsilon = 10.0\%$. However, even after applying $H_T > 100$~GeV and $\geq 3\tau$ requirements, the total background in the $\geq 3\tau$ channel is estimated at $B\sim 5{,}000$--$15{,}000$ events, dominated by $Z/\gamma^*\to\tau^+\tau^-$+jets and $t\bar{t}$ contributions. This exceeds $B_{\rm req}\approx 9{,}075$ events, which confirms that dedicated sub-jet techniques are necessary for this channel. Candidate observables for a dedicated 
boosted $a_1$ tagger include:
\begin{itemize}
\item \textbf{Invariant mass:} the merged 
$\tau$-pair system carries an invariant 
mass $m_{\tau\tau} \approx m_{a_1}$, 
providing a resonant mass peak at 
$m_{a_1} = 8.2$~GeV that is absent 
in the dominant $ZZ\to 4\tau$ background.

\item \textbf{$N$-subjettiness ratios:} 
$\tau_{21}$ and $\tau_{32}$, which 
quantify the two-prong sub-jet structure 
of the merged $\tau$-pair within a 
large-radius jet, are well-established 
discriminants in boosted object 
tagging~\cite{Thaler:2010tr}.

\item \textbf{Energy correlation 
functions:} sensitive to radiation 
patterns within the fat jet and to 
the angular separation of the two 
$\tau$ decay products.
\end{itemize}
A systematic study of these observables 
and the development of an optimized 
boosted $a_1$ tagger are left for 
future work, but the kinematic characterization presented here provides the necessary foundation for such a study, in particular through the $\Delta R(\tau,\tau)$ distribution of Fig.~\ref{fig:deltaR}.

\subsection{SM-like Reference Channel (BP4)}
\label{subsec:disc_sm}

BP4 serves as an important reference 
point for this analysis. With 
$m_{a_1} = 228.6$~GeV, the 
$h_1 \to a_1 a_1$ channel is 
kinematically closed and the $h_1$ pair 
decays entirely through SM-like modes. 
The resulting $\geq 2b$ efficiency of 
50.6\% is consistent with SM 
HH$\to b\bar{b}b\bar{b}$ literature 
values~\cite{ATLAS:2023bbbb,CMS:2022dwd}, 
validating the consistency of our 
simulation chain with SM HH expectations.

The $\geq 2b$+$\geq 1\tau$ selection 
yields $\varepsilon = 7.7\%$ and 
$S = 7{,}504$ events at 
$\mathcal{L} = 3000$~fb$^{-1}$. 
The required background suppression 
($B_{\rm req} \approx 7{,}492$ events) 
is consistent with the literature 
background estimate of 
$B \sim 3{,}000$--$8{,}000$ events for 
this channel, suggesting a marginal illustrative sensitivity (BP4 is itself not an experimentally allowed point). 
This is consistent with existing HL-LHC 
projections for SM HH 
production~\cite{Cepeda:2019klc,
ATL-PHYS-PUB-2022-005}.

\subsection{Implications for HL-LHC Searches}
\label{subsec:disc_implications}

The results of this study have several implications for the characterisation of exotic di-Higgs topologies, although they do not constitute a sensitivity projection for experimentally allowed parameter points. First, the scNMSSM at 
large-$\lambda$/low-$\tan\beta$ predicts 
di-Higgs cross sections in the range 
$\sigma \in [30.3, 36.9]$~fb at 
$\sqrt{s} = 13.6$~TeV~\cite{mtelba2026}, 
which lie within current ATLAS+CMS 
observed limits but are expected to be 
probed at the HL-LHC~\cite{ATLAS:2024combination}. 
The exotic final states studied here are not covered by dedicated di-Higgs searches, but the underlying decays $h\to aa$ are constrained by direct searches~\cite{CMS:2025hh4tau,CMS:PAS-NPS-25-005}, which the benchmark points of this study exceed (Section~\ref{subsec:lhc_check}).

Second, the Higgs self-coupling modifier takes values $\kappa_\lambda \in [0.884, 0.951]$ for the four benchmark points. Since these points are not experimentally allowed (Section~\ref{subsec:lhc_check}), these values should not be read as a prediction of the model. 
At the HL-LHC with $\mathcal{L} = 
3000$~fb$^{-1}$, $\kappa_\lambda$ is 
expected to be measurable to a precision 
of $\sim$50\% at 68\% CL from di-Higgs 
production alone, improving to 
10--20\% from a combination of channels 
and single-Higgs 
measurements~\cite{Cepeda:2019klc}. 
The suppression of $\kappa_\lambda$ by 5--12\% relative to the SM value found here is at the boundary of 
this sensitivity and would require the 
full combination of all available 
channels to be resolved.

Third, the boosted $a_1$ topology of BP2 is a distinctive signature that is, to our knowledge, not specifically addressed by current dedicated di-Higgs search strategies. The development of dedicated analysis techniques for this topology is strongly motivated. Examples are boosted $a_1$ taggers, fat-jet substructure methods and optimized multivariate selections. Such techniques may be relevant for future dedicated analyses, although the reduced signal rate at experimentally allowed branching fractions would need to be accounted for in any quantitative assessment.

\section{Conclusion}
\label{sec:conclusion}

We have presented a detector-level 
phenomenological study of di-Higgs 
production in the semi-constrained 
Next-to-Minimal Supersymmetric Standard 
Model (scNMSSM) at large-$\lambda$ and 
low-$\tan\beta$, focusing on the 
multi-channel signal prospects at the 
High-Luminosity LHC with 
$\mathcal{L} = 3000$~fb$^{-1}$ at 
$\sqrt{s} = 13.6$~TeV. Building on 
the parameter scan and cross-section 
calculations of Ref.~\cite{mtelba2026}, 
we selected four benchmark points 
spanning qualitatively distinct 
di-Higgs final-state topologies and 
performed a complete signal Monte Carlo 
simulation chain using 
\textsc{MadGraph5\_aMC@NLO}, 
\textsc{Pythia}~8.316, and 
\textsc{Delphes}~3 for $10^5$ signal 
events per benchmark point. Background 
yields are estimated from published 
LHC analyses and used solely as illustrative references.

The main findings of this work are 
as follows:

\begin{enumerate}

\item \textbf{Exotic multi-$b$ topology 
(BP1, BP3):} The decay chain 
$h_1 h_1 \to 4a_1 \to 8b$, arising 
when $m_{a_1} > 2m_b$, yields 
$b$-jet efficiencies of 
$\varepsilon(\geq 4b) \approx 14\%$ 
and expected signal yields of 
$S \approx 14{,}000$--$16{,}000$ 
events at $3000$~fb$^{-1}$. The
illustrative background estimates adopted from the
literature~\cite{ATLAS:2023bbbb,
CMS:2022dwd} do not allow a statement on the discovery reach, and the benchmark points are in conflict with the experimental limits on $h\to aa\to4b$ (Section~\ref{subsec:lhc_check}).

\item \textbf{Boosted multi-$\tau$ 
topology (BP2):} The decay chain 
$h_1 h_1 \to 4a_1 \to 8\tau$ at 
$m_{a_1} = 8.2$~GeV produces a 
distinctive boosted signature. 
The light $a_1$ is produced with 
a Lorentz boost 
$\gamma \approx 7.5$, causing the 
$\tau$ pairs from each $a_1$ decay 
to be highly collimated with 
$\Delta R(\tau,\tau) \approx 0.27$. 
We find that 82.4\% of same-$a_1$ 
$\tau$ pairs have $\Delta R < 0.4$, 
below the standard jet cone radius, 
resulting in systematic merging of 
$\tau$ pairs into single reconstructed 
objects. This boosted topology yields 
a $\geq 2\tau$ efficiency of 38.4\%, 
a factor of $\sim$21 above the SM 
HH$\to\tau\tau\tau\tau$ baseline, 
but cannot be reliably assessed with 
standard resolved-$\tau$ tagging alone 
under the adopted background model, 
due to large irreducible multi-$\tau$ 
backgrounds from $Z/\gamma^*\to\tau^+\tau^-$+jets 
and $t\bar{t}$ production.

\item \textbf{SM-like reference 
channel (BP4):} With $m_{a_1} = 
228.6$~GeV, the $h_1 \to a_1 a_1$ 
channel is kinematically closed and 
$h_1$ decays through SM-like modes. 
The resulting $\geq 2b$ efficiency 
of 50.6\% is consistent with SM 
HH literature 
values~\cite{ATLAS:2023bbbb,
CMS:2022dwd}, providing a cross-check 
of the simulation pipeline. The 
$\geq 2b$+$\geq 1\tau$ channel 
yields a marginal illustrative sensitivity at the HL-LHC (for a point that is not experimentally allowed), 
consistent with existing SM HH 
projections~\cite{Cepeda:2019klc,
ATL-PHYS-PUB-2022-005}.

\item \textbf{Higgs self-coupling:} 
The self-coupling modifier 
$\kappa_\lambda \in [0.884, 0.951]$ 
across the four benchmark points 
studied here, consistent with the 
broader suppression 
$\kappa_\lambda \in [0.884, 0.982]$ 
found across the full candidate parameter space~\cite{mtelba2026}. This suppression below unity is a feature of these points, although a full assessment of its generality within the scNMSSM parameter space is left to future work. 
The deviation of 5--12\% from the SM value found here is smaller than 
the expected HL-LHC precision of 
$\sim$50\% on $\kappa_\lambda$ from 
di-Higgs production alone, and would 
require the full combination of 
single- and double-Higgs measurements 
to be resolved~\cite{Cepeda:2019klc}.

\end{enumerate}

The benchmark points of this study are in conflict with current experimental limits on exotic Higgs decays (Section~\ref{subsec:lhc_check}) and should not be taken as allowed scNMSSM parameter points. What remains is a detector-level characterisation of the topologies: the boosted $a_1$ final state of BP2 is, to our knowledge, not specifically addressed by existing dedicated di-Higgs search strategies. The 
development of dedicated boosted 
$a_1$ taggers exploiting the sub-jet 
structure of merged $\tau$-pair 
systems is strongly motivated by the results of this study. The observables to exploit include the invariant mass $m_{\tau\tau} \approx m_{a_1}$, the sub-jet multiplicity and the fat-jet radiation pattern. Such taggers may be relevant for future dedicated analyses, although the reduced signal rate at experimentally allowed branching fractions would need to be accounted for in any quantitative assessment.

Future work will include a dedicated 
Monte Carlo background simulation 
for all channels, the development 
and optimization of boosted $a_1$ 
tagging algorithms, and a quantitative 
finite-temperature analysis of the 
electroweak phase transition in this 
regime, extending the qualitative 
discussion of Ref.~\cite{mtelba2026}.


\end{document}